\documentclass[twocolumn]{aastex631}
\usepackage{graphicx}
\usepackage{mathtools}
\graphicspath{{./figures/}}
\usepackage{amsmath}
\usepackage[utf8]{inputenc}
\usepackage[OT6,T1]{fontenc}

\newcommand{\Msolar}{M$_{\odot}$}
\newcommand{\Rsolar}{R$_{\odot}$}

\newcommand{\logg}{log$_{10}\left(g\right)$}
\newcommand{\teff}{$T_\mathrm{eff}$}
\newcommand{\vsini}{$v\,$sin$\,i$}
\newcommand{\halpha}{H$\alpha$}

\shorttitle{Detection of Hot WD Companions to Blue Lurkers}
\shortauthors{Nine et al.}

\begin{document}

\title{WIYN Open Cluster Study. LXXXVII. \textit{HST} Ultraviolet Detection of Hot White Dwarf Companions to Blue Lurkers in M67}

\author[0000-0002-6478-0611]{Andrew C. Nine}
\affiliation{Department of Astronomy, University of Wisconsin-Madison, 475 N Charter St., Madison, WI 53706, USA}
\email{anine@astro.wisc.edu}

\author[0000-0002-8308-1998]{Robert D. Mathieu}
\affiliation{Department of Astronomy, University of Wisconsin-Madison, 475 N Charter St., Madison, WI 53706, USA}

\author[0000-0002-8443-0723]{Natalie M. Gosnell}
\affiliation{Department of Physics, Colorado College, 14 East Cache La Poudre Street, Colorado Springs, CO 80903, USA}

\author[0000-0002-3944-8406]{Emily M. Leiner}
\altaffiliation{NSF Astronomy \& Astrophysics Fellow}
\affiliation{Center for Interdisciplinary Exploration and Research in Astrophysics, Northwestern University,\\ 2145 Sheridan Rd., Evanston, IL 60208, USA}

\begin{abstract}
    We present the results of our \textit{Hubble Space Telescope} far-ultraviolet survey of the blue lurkers (BLs) in M67. We find evidence for two white dwarf companions among the BLs that are indicative of mass transfer from an evolved companion, one in WOCS 14020 and the other in WOCS 3001. The cooling ages of the white dwarfs suggest that mass transfer in these systems occurred $\sim$300--540 Myr and $\sim$600--900 Myr ago, respectively. The rotation periods and cooling ages of the BLs are consistent with spin-up and subsequent single-star spin-down models, and binary evolution models yield plausible evolutionary pathways to both BLs via highly non-conservative mass transfer. We conclude that the BLs are lower-luminosity analogues to the classical blue stragglers.
    \vspace{0.75cm}
\end{abstract}

\section{Introduction}
\label{sec:intro}

Open clusters provide a rich environment in which to study the interplay between binarity and stellar evolution. Binary interactions can create stars that challenge the standard model of single-star evolution. The most well-known of these alternative stellar evolution products are the blue straggler stars (BSSs), first observed by \cite{Sandage1953} as stars that were brighter and bluer than the main-sequence (MS) turnoff in a color-magnitude diagram (CMD). There have been several other types of stars discovered in open clusters that do not lie on single-star evolutionary tracks, including yellow stragglers (\citealt{Strom1971, Landsman1997, Leiner2016}), sub-subgiants (\citealt{Mathieu2003, Geller2017, Gosnell2022}), and red stragglers (\citealt{Geller2017}). BSSs and other alternative stellar evolution products comprise 25\% of the evolved stars in older open clusters such as M67 (4 Gyr; \citealt{MathieuLeiner2019}).

Current theories for BSS formation rest on the overarching idea that they are MS stars that have recently gained mass. Three channels are most frequently considered: mergers of a binary through such means as wind mass loss or the Kozai mechanism (\citealt{Andronov2006, Perets2009, Andrews2016}), the collision of two stars within a dynamical encounter (\citealt{Hills1976, PZ2010}), and mass transfer from a companion star (\citealt{McCrea1964, Chen2008, Mathieu2009}).

Studies of the old open cluster NGC 188 (7~Gyr) found that approximately 75\% of 21 BSSs were spectroscopic binary stars (\citealt{Mathieu2009}). Their secondary mass distribution showed a sharp peak at 0.5~\Msolar, consistent with the presence of CO white dwarfs (WDs; \citealt{Geller2011}). Further observations with the \textit{Hubble Space Telescope} (\textit{HST}) detected four hot ($>$13,000 K) WD companions to these BSSs and found evidence for three additional cooler ($>$10,000 K) WD companions (\citealt{Gosnell2015}). The masses of two of the hot WD companions were later measured through far-ultraviolet (FUV) spectroscopy (\citealt{Gosnell2019}). These results indicate that mass transfer is the dominant mechanism for the formation of the BSSs in NGC 188.

Recently, a new alternative stellar evolution product was discovered in M67: the blue lurkers (BLs). First described in \citet[hereafter L19]{Leiner2019}, the BLs are stars within the MS of M67 that are anomalously rapidly rotating given their age, with rotation periods ranging between 2--8 days. Being within the single-star MS, the BLs do not stand out in an optical CMD as do the other known alternative stellar evolution products. Of the eleven BLs described in L19, eight are in single-lined spectroscopic binary systems with periods between $10^2$ and $10^4$ days, too wide for the primary stars to have been spun up by tidal interactions. L19 hypothesized that the BLs are the result of mass transfer from an evolved companion or of a merger or collision between two MS stars, and are therefore low-luminosity analogues of the classical BSSs. \cite{Jadhav2019} subsequently detected a significant excess in FUV flux from the BL WOCS 3001 (see Section \ref{subsec:wd_detections}), which they interpret to be from a hot WD companion. A further three BL candidates were identified in Ruprecht 147 (Ru 147; 2.7~Gyr) by \cite{Curtis2020}, two of which have similar masses to the BLs of M67. The BL candidates in Ru 147 were not observed to be velocity-variable, which \cite{Curtis2020} hypothesized may have been the result of mergers or having binary companions with orbits oriented along the plane of the sky.

We set out to search for WD companions to the BLs of M67, which would indicate that they were formed by mass transfer. If found, we further seek to connect the formation history of the BLs to the classical BSSs and other mass-transfer products.

In this work we present the results of our \textit{HST} study of the binary BLs in M67 conducted as part of the ongoing WIYN Open Cluster Study (WOCS; \citealt{Mathieu2000}), and we discuss their implications for possible BL formation mechanisms. We describe our observations in Section \ref{sec:obs}, and our photometric analysis of the BLs and the detection of two hot WD companions in Section \ref{sec:analysis}. We discuss the properties of the BL population and possible evolutionary scenarios in Section \ref{sec:discuss}, and present our summary in Section \ref{sec:summary}.

\section{Observations}
\label{sec:obs}

\begin{figure}[t]
    \centering
    \includegraphics[width=\linewidth]{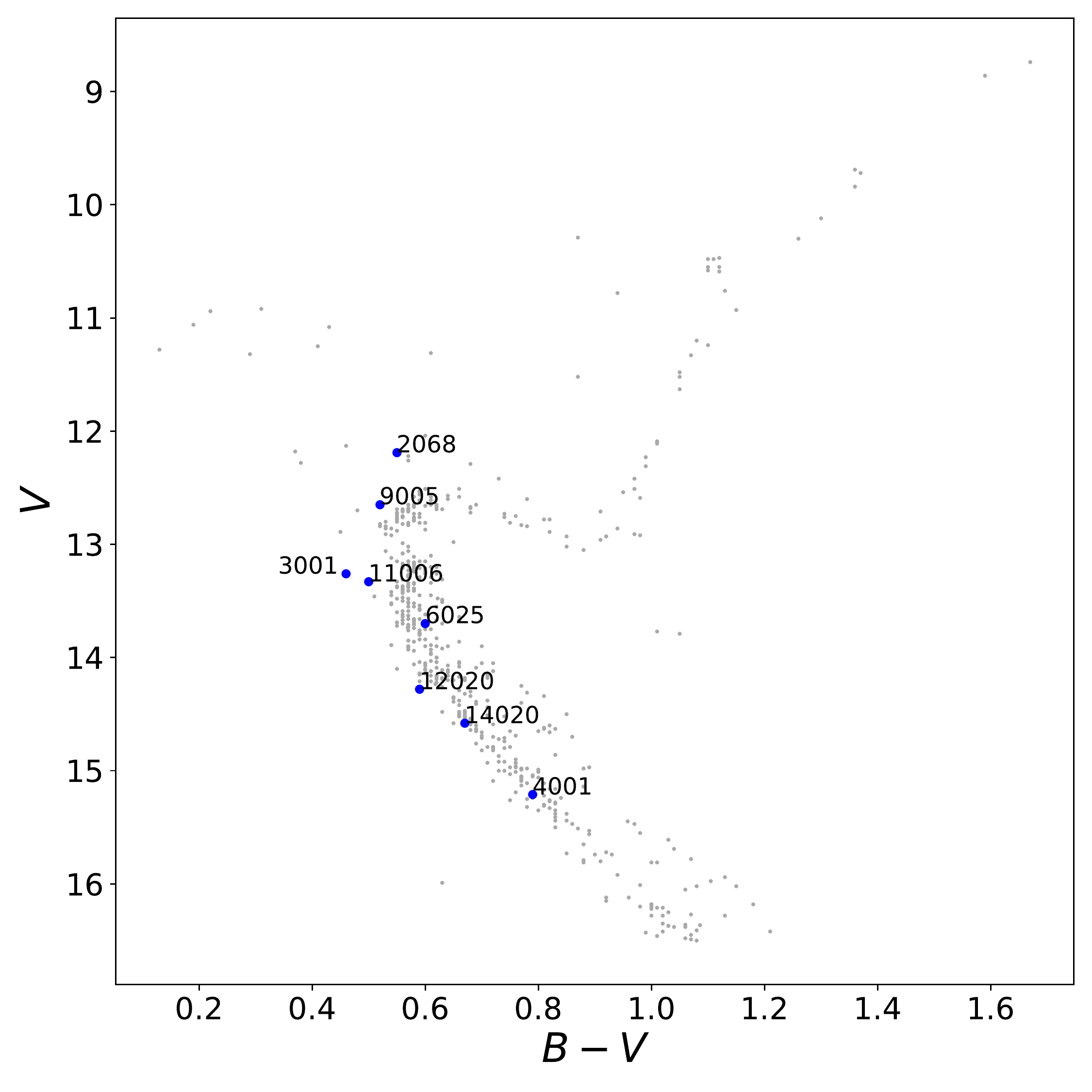}
    \caption{Optical CMD of M67, with member stars from \cite{Geller2021} plotted as gray points. We mark the eight binary BLs with blue dots and label them with their corresponding WOCS IDs.}
    \label{fig:bl_cmd}
\end{figure}

We provide a summary of our BL sample in Table \ref{blue:tab:summ}, including their J2000 positions, optical photometry and colors, effective temperatures, orbital parameters, and rotation periods. We also highlight the eight observed BLs in an optical CMD for reference (Figure \ref{fig:bl_cmd}). The complete population is described in L19.

We follow a similar observational design as detailed in \cite{Gosnell2014} and \cite{Gosnell2015}. We observed the eight binary BLs in the FUV using the \textit{HST} Advanced Camera for Surveys (ACS) in the Solar Blind Channel (SBC). The observations took place over 17 orbits between 1 March 2021 and 26 April 2021 (GO: 16244, PI: Mathieu), with one re-observation of WOCS 2068 occurring on 4 January 2022 due to poor tracking. We also observed a red giant member of M67, WOCS 1003 ($V=10.43$, $B-V=1.11$; \citealt{Montgomery1993}), in order to determine the magnitude of the ACS red leak (see Section \ref{subsec:photo}).  

Each of the BLs in our sample were observed in F140LP for a total of 1800~s, F150LP for 2160~s, and F165LP for 1174~s. We also take advantage of the nested structure of these three filters in order to derive narrow bandpasses that better isolate the FUV fluxes of these BLs. For each observed star, we take the difference in the observed count rates between F140LP and F150LP in order to derive the flux in the F140N bandpass, and between F150LP and F165LP to derive the flux in the F150N bandpass. The throughput curves of these bandpasses are shown for reference in Figure \ref{fig:bl_sed_14020}; see also \cite{Dieball2005} and \cite{Gosnell2015}.

\subsection{Aperture Photometry}
\label{subsec:photo}

We carry out aperture photometry on the redrizzled images with a pixel scale of 0.025$''$ per pixel. We extract count rates using an aperture radius of 40 pixels, or 1$''$, using \texttt{photutils}\footnote{\url{https://photutils.readthedocs.io/en/stable/}} (\citealt{Bradley2021}). We calculate the encircled energy corrections following the results of \cite{Avila2016}, who found the encircled energy correction factor to be approximately 0.92 at a radius of 1$''$ in each of F140LP, F150LP, and F165LP.

\movetabledown=1.2in
\centerwidetable
\begin{rotatetable}
\begin{deluxetable*}{ccccccccccccccc}
\tabletypesize{\scriptsize}
\tablecaption{Blue Lurker Summary Table\label{blue:tab:summ}}
\tablehead{\colhead{WOCS ID} & \colhead{Cross ID} & \colhead{$\alpha$ (J2000)} & \colhead{$\delta$ (J2000)} & \colhead{$V$\tablenotemark{a}} & \colhead{$B-V$\tablenotemark{a}} & \colhead{$T_{\mathrm{eff}}$, SED\tablenotemark{b}} & \colhead{$T_{\mathrm{eff}}$, H $\alpha$\tablenotemark{b}} & \colhead{$T_{\mathrm{eff}}$, GALAH\tablenotemark{c}} & \colhead{\logg\tablenotemark{d}} & \colhead{P$_{\mathrm{orb}}$\tablenotemark{e}} & \colhead{Eccentricity\tablenotemark{e}} & \colhead{$f(m)$\tablenotemark{e}} & \colhead{\vsini} & \colhead{P$_{\mathrm{rot}}$\tablenotemark{f}} \\
\colhead{} & \colhead{\cite{Sanders1977}} & \colhead{} & \colhead{} & \colhead{} & \colhead{} & \colhead{(K)} & \colhead{(K)} & \colhead{(K)} & \colhead{} & \colhead{(days)} & \colhead{} & \colhead{(\Msolar)} & \colhead{(km sec$^{-1}$)} & \colhead{(days)}}
\tablewidth{0pt}
\startdata
4001 & 1029 & 8 51 21.62 & 11 49 02.5 & 15.21 & 0.79 & $5580^{+60}_{-130}$ & \nodata & $5400 \pm 90$ & \nodata & 139.77 & 0.36 & 2.17e-2 & <10 & 7.9 \\
14020 & 1452 & 8 52 03.50 & 11 47 48.1 & 14.58 & 0.67 & $5990^{+60}_{-110}$ & $5900^{+130}_{-150}$ & $5790 \pm 80$ & $4.45^{+0.03}_{-0.04}$ & 358.9 & 0.23 & 2.38e-3 & <10 & 4.4 \\
12020 & 1102 & 8 51 08.85 & 11 57 53.7 & 14.28 & 0.59 & $6190^{+100}_{-140}$ & $5960^{+100}_{-110}$ & $5920 \pm 130$ & $4.44^{+0.02}_{-0.03}$ & 762 & 0.056 & 2.87e-2 & <10 & 7.6 \\
3001 & 1031 & 8 51 22.96 & 11 49 13.1 & 13.26 & 0.46 & $6690^{+80}_{-160}$ & $6430^{+70}_{-80}$ & $6420 \pm 80$ & $4.29^{+0.03}_{-0.03}$ & 128.14 & 0.04 & 1.43e-2 & $24.7\pm0.6$ & 2.0$^{+0.6}_{-0.9}$ \\
2068 & 277 & 8 49 21.48 & 12 04 22.8 & 12.19 & 0.55 & 6000, 6800 & \nodata & \nodata & \nodata & 8567 & 0.859 & 6.81e-2 & \nodata & 20 and 3 \\
9005 & 1005 & 8 51 15.45 & 11 47 31.4 & 12.65 & 0.52 & $6500^{+90}_{-110}$ & $6240^{+100}_{-90}$ & $6290 \pm 80$ & $3.98^{+0.03}_{-0.03}$ & 2769 & 0.15 & 3.68e-2 & <10 & 4.5 \\
6025 & 1431 & 8 52 05.81 & 11 42 24.7 & 13.70 & 0.60 & $6200^{+150}_{-230}$ & $6010^{+80}_{-100}$ & \nodata & $4.23^{+0.03}_{-0.03}$ & 6265 & 0.38 & 2.20e-1 & <10 & 2.3 \\
11006 & 2223 & 8 51 29.86 & 11 51 29.9 & 13.33 & 0.50 & $6540^{+210}_{-200}$ & \nodata & $6370 \pm 110$ & \nodata & $>3500$ & \nodata & \nodata & $27.9^{+0.8}_{-0.9}$ & 1.8$^{+0.5}_{-0.8}$ \\
\enddata
\tablenotetext{a}{\cite{Montgomery1993}}
\tablenotetext{b}{Computed for this work (Section \ref{subsec:blmods}), except for WOCS 2068. See Section \ref{subsec:photo}.}
\tablenotetext{c}{DR3; \cite{Buder2021}}
\tablenotetext{d}{Estimated with \texttt{isochrones} (\citealt{Morton2015}). See Section \ref{subsec:blmods}.}
\tablenotetext{e}{\cite{Geller2021}}
\tablenotetext{f}{\cite{Leiner2019}, except for WOCS 3001 and WOCS 11006. See Section \ref{subsec:blmods}}
\end{deluxetable*}
\end{rotatetable}
\clearpage

In the case of WOCS 2068, we find a nearby non-member star within the 1$''$ aperture, \textit{Gaia} DR2 604984665203481344 ($G=14.1$; \citealt{GAIA2016, GaiaCollab2018a, GaiaCollab2018b}). This source is not listed in the \textit{Gaia} DR3 catalog (\citealt{GaiaDR3}). In order to correct for this, we computed the count rate for the secondary source in a 0.5$''$ aperture in each \textit{HST} filter. We then applied the appropriate encircled energy and background corrections following \cite{Avila2016}, and subtracted the determined count rates from the respective measurements for WOCS 2068. The corrections obtained in this process are of order 0.1 mag, but the added noise contribution from the secondary source reduces the detection significance of our measured FUV magnitudes for WOCS 2068. The corrected FUV magnitudes of WOCS 2068 are well-fit by the composite stellar SED of the BSS and a MS companion proposed by L19 when convolved with the \textit{HST} FUV bandpasses with \texttt{pysynphot}\footnote{\url{https://pysynphot.readthedocs.io/en/latest/}} (\citealt{Lim2015}).

The ACS/SBC has a known red leak beyond 2000 \r{A} (\citealt{Weaver2010}). In order to correct for this red leak, we follow a similar procedure as \cite{Gosnell2015}. We assume that the observed FUV count rate of the red giant WOCS 1003 is due entirely to the red leak. Our measured count rates for WOCS 1003 are $2.96 \pm 0.04$, $2.91 \pm 0.04$, and $2.91 \pm 0.05$ counts~s$^{-1}$ in F140LP, F150LP, and F165LP, respectively. For comparison, the expected count rates for WOCS 1003 if there were no red leak in ACS/SBC are approximately 0.002 counts~s$^{-1}$ in each filter, as computed with \texttt{pysynphot}. We scale the observed count rates from WOCS 1003 by the apparent $V$ magnitude of each of the BLs in our sample. The scaled red leak count rates are then subtracted off from the observed count rates of the BLs in each filter. Typical magnitude corrections obtained in this process are of order 0.1 mag. 

\subsection{Magnitude Calculation}
\label{subsec:mags}

We calculate the instrumental magnitudes for each BL by first deriving the instrumental zeropoint in each filter through the relation 

\begin{equation*}
    \text{ZPT} = -2.5 \log_{10}(\text{PHOTFLAM}) - 21.1,
\end{equation*}

\noindent where PHOTFLAM is the inverse sensitivity obtained from the image headers, and then converting the corrected count rates to the STMAG system through

\begin{equation*}
    \text{STMAG} = -2.5 \log_{10}(\text{count rate})+\text{ZPT}.
\end{equation*}

\noindent These formulae are described in \cite{Lucas2021}. The inverse sensitivities for F140N and F150N are computed with \texttt{pysynphot}, with the results as follows:

\begin{equation*}
    \text{PHOTFLAM}_{\text{F140N}} = 5.5079\times10^{-17}\, \text{erg}\, \text{cm}^{-2}\, \text{\r{A}}^{-1}\, \text{count}^{-1},
\end{equation*}
\begin{equation*}
    \text{PHOTFLAM}_{\text{F150N}} = 5.0702\times10^{-17}\, \text{erg}\, \text{cm}^{-2}\, \text{\r{A}}^{-1}\, \text{count}^{-1}.
\end{equation*}

\noindent We treat the photometric errors in the broadband filters as Poisson-dominated, and we compute the errors in the derived narrow bands by adding the contributions to the Poisson noise from the component filters in quadrature.

We present the FUV measurements for the 8 BLs in Table \ref{blue:tab:uvtab}. We only give a magnitude in F140N for those sources with a higher count rate in F140LP than in F150LP, and likewise for F150N. Magnitudes in italics are less than 3$\sigma$ detections compared to the observational noise. 

\centerwidetable
\begin{deluxetable*}{ccccccccc}
\tabletypesize{\footnotesize}
\tablecaption{Blue Lurker FUV Photometry}\label{blue:tab:uvtab}
\tablehead{\colhead{WOCS ID} & \colhead{FUV\tablenotemark{a}} & \colhead{NUV\tablenotemark{a}} & \colhead{F140LP} & \colhead{F150LP} & \colhead{F165LP} & \colhead{F140N\tablenotemark{b}} & \colhead{F150N\tablenotemark{b}}}
\tablewidth{0pt}
\startdata
4001 & \nodata & \nodata & 24.06 $\pm$ 0.21 & 23.28 $\pm$ 0.16 & 21.72 $\pm$ 0.19 & \nodata & \nodata \\
14020 & 21.5 $\pm$ 0.2 & 19.55 $\pm$ 0.05 & 19.04 $\pm$ 0.01 & 18.93 $\pm$ 0.01 & 18.78 $\pm$ 0.03 & 19.24 $\pm$ 0.05 & 18.90 $\pm$ 0.03 \\
12020 & 23.9 $\pm$ 0.4 & 19.11 $\pm$ 0.01 & 21.04 $\pm$ 0.03 & 20.44 $\pm$ 0.03 & 19.22 $\pm$ 0.05 & \nodata & \textit{22.5}$^{+\textit{0.7}}_{-\textit{0.4}}$ \\
3001 & \nodata & 17.21 $\pm$ 0.03 & 19.42 $\pm$ 0.02 & 18.84 $\pm$ 0.01 & 17.66 $\pm$ 0.02 & \nodata & 20.49$^{+0.18}_{-0.15}$ \\
2068 & 22.8 $\pm$ 0.2 & 16.72 $\pm$ 0.01 & 18.86 $\pm$ 0.01 & 18.39 $\pm$ 0.01 & 17.17 $\pm$ 0.02 & \textit{20.9}$\pm$\textit{0.3} & 20.36$^{+0.21}_{-0.17}$ \\
9005 & \nodata & 16.91 $\pm$ 0.02 & 19.00 $\pm$ 0.01 & 18.54 $\pm$ 0.01 & 17.28 $\pm$ 0.02 & 20.9$^{+0.3}_{-0.2}$ & 20.9 $\pm$ 0.3 \\
6025 & 24.0 $\pm$ 0.4 & 18.43 $\pm$ 0.02 & 20.83 $\pm$ 0.03 & 20.37 $\pm$ 0.03 & 19.09 $\pm$ 0.05 & \textit{22.7}$^{+\textit{1.0}}_{-\textit{0.5}}$ & \textit{23.1}$^{+\textit{2.4}}_{-\textit{0.7}}$ \\
11006 & 21.8 $\pm$ 0.2 & 16.79 $\pm$ 0.02 & 19.42 $\pm$ 0.02 & 18.92 $\pm$ 0.01 & 17.68 $\pm$ 0.02 & \textit{22.0}$^{+\textit{0.9}}_{-\textit{0.5}}$ & 21.1 $\pm$ 0.3 \\
\enddata
\tablenotetext{a}{\textit{GALEX} measurements (AB system; \citealt{Martin2005}).}
\tablenotetext{b}{Magnitudes in italics are less than 3$\sigma$ detections.}
\end{deluxetable*}

\section{Analysis}
\label{sec:analysis}

\subsection{The Blue Lurker Effective Temperatures}
\label{subsec:blmods}

Identifying hot WD companions to the BLs relies on the detection of FUV flux significantly above what is expected for a single BL given its effective temperature. In order to constrain the effective temperature of the BL primary stars, and therefore their photospheric FUV flux, we first employ a similar SED fitting technique as described in L19. Because all of the binaries are single-lined, there is no need to consider contamination of the SED from a secondary non-WD companion, with the exception of WOCS 2068 for which there is an indication in the SED of a MS stellar companion (see Section \ref{subsec:photo}).

We first obtain published photometric measurements for each of the BLs through the SIMBAD database (\citealt{Wenger2000}). We compile IR photometry from 2MASS (\citealt{Cutri2003, Skrutskie2006}) and WISE (\citealt{Wright2010}), and optical photometry from \cite{Montgomery1993}, \cite{Ducourant2006}, \cite{Pasquini2008}, \cite{KroneMartins2010}, \cite{Pace2012}, and  \cite{Munari2014}, as available. We then deredden each photometric measurement assuming $E(B-V) = 0.041 \pm 0.004$ (\citealt{Taylor2007}) and using the extinction curve of \cite{Cardelli1989}. We then generate a grid of \cite{Castelli2003} model atmospheres using \texttt{pysynphot}, assuming
solar metallicity and normalizing each model atmosphere to the dereddened $V$ magnitude. We then increment the effective temperature of the model atmospheres in steps of 10 K until we find the best-fit model through $\chi^2$ minimization. We implement a bootstrap routine that resamples the available photometric measurements and refits the model atmospheres in order to derive uncertainties on the effective temperature. Our derived effective temperatures obtained in this method match those found by L19 within our uncertainties.

As an alternative approach, we have also obtained \halpha\, spectra for five of the BLs with the Hydra Multi-Object Spectrograph (MOS; \citealt{Barden1994}) on the WIYN\footnote{The WIYN 3.5m Observatory is a joint facility of the University of Wisconsin-Madison, Indiana
University, and the National Optical-Infrared Astronomy Research Laboratory.} 3.5m telescope, which we use to further constrain their effective temperatures. The spectra have a typical signal-to-noise ratio of $>100$ and $R\sim17,000$. Detailed analysis of the complete spectra will be the subject of future work. 

In order to derive the effective temperatures from the \halpha\, profiles, we first compute continuum-normalized spectra with \texttt{specutils}\footnote{\url{https://specutils.readthedocs.io/en/stable/index.html}} (\citealt{Earl2022}). We then create a grid of theoretical solar-metallicity \halpha\, profiles\footnote{The \cite{Barklem2002} profiles are publicly available at \url{https://github.com/barklem/public-data}.} obtained from \cite{Barklem2002}, spanning \logg~=~3.6--4.5 and \teff~=~5000--7500 K. We estimate the values of \logg~ for each BL with the \texttt{starfit} function of the \texttt{isochrones}\footnote{\url{https://github.com/timothydmorton/isochrones}} package (\citealt{Morton2015}), which estimates stellar parameters through interpolation across a grid of MIST isochrones (\citealt{Dotter2016, Choi2016}). We include the estimated values of \logg\, in Table \ref{blue:tab:summ}.

In order to find the best-fit model profile to each \halpha\, observation, we determine the \vsini~ of each BL following the method of \cite{Rhode2001}. Our minimum velocity resolution with the Hydra MOS is $\sim$10~km~sec$^{-1}$ (\citealt{Rhode2001, Geller2009}); we therefore do not report \vsini~ values less than this limit in Table \ref{blue:tab:summ}. We find, as did L19, that both WOCS 3001 and WOCS 11006 are rotating more rapidly than this limit, though our derived \vsini~ values are $\approx$ 50\% higher than L19. We report the updated values in Table \ref{blue:tab:summ} along with their corresponding rotation periods, which we derive in a similar manner as L19. We then interpolate across the grid of \cite{Barklem2002} profiles, which we rotationally broaden to match the observed \vsini~ of each BL using the convolution method of \cite{Gray2005}, as implemented in \texttt{PyAstronomy}\footnote{\url{https://github.com/sczesla/PyAstronomy}} (\citealt{Czesla2019}). We find the best-fit \teff~ and uncertainties for each BL through $\chi^2$ minimization from the fit of the wings of the broadened profiles to the observed spectra, which are listed in Table \ref{blue:tab:summ}. For comparison, we also include in Table \ref{blue:tab:summ} the spectroscopic temperatures from GALAH DR3 (\citealt{Buder2021}).

We note that there is a systematic difference of $\sim$200~K between our SED temperatures and our \halpha\, temperatures, in the sense of lower \halpha\, temperatures. Similar offsets have been observed previously when comparing photometric and spectroscopic methods (\citealt{Escorza2017, Shetye2021, Buder2021}). \halpha\, profile fitting does not depend on reddening and is only weakly dependent on other stellar parameters such as \logg~  and metallicity (\citealt{Fuhrmann1994, Barklem2002, Giribaldi2019}), making it a well-suited method to determine the effective temperatures of FGK-type stars. We therefore adopt our \halpha-derived temperatures for the remainder of our analysis.

\subsection{The M67 Model Population}
\label{subsec:msmods}

Since the BLs lie within or near the MS of an optical CMD, we compare the expected FUV flux of M67 MS and evolved stars to the observed FUV flux from the BLs. In order to model the underlying stellar population of M67, we use all of the known member stars from \cite{Geller2021}, excluding the BSSs. We estimate the effective temperatures of the member stars from their published photometry using the relations of \cite{Ramirez2005} in the interest of computational efficiency. We then generate UVBLUE\footnote{\url{https://www.inaoep.mx/~modelos/uvblue/uvblue.html}} (\citealt{RM2005}) model atmospheres corresponding to each computed effective temperature, normalize them to their expected \textit{GALEX} NUV magnitudes based on their published $V$ magnitudes, and convolve them with the \textit{HST} FUV bandpasses with \texttt{pysynphot}. We also calculate the expected FUV narrowband colors in order to compare with the BL observations. 

\begin{figure}[t]
    \centering
    \includegraphics[width=\linewidth]{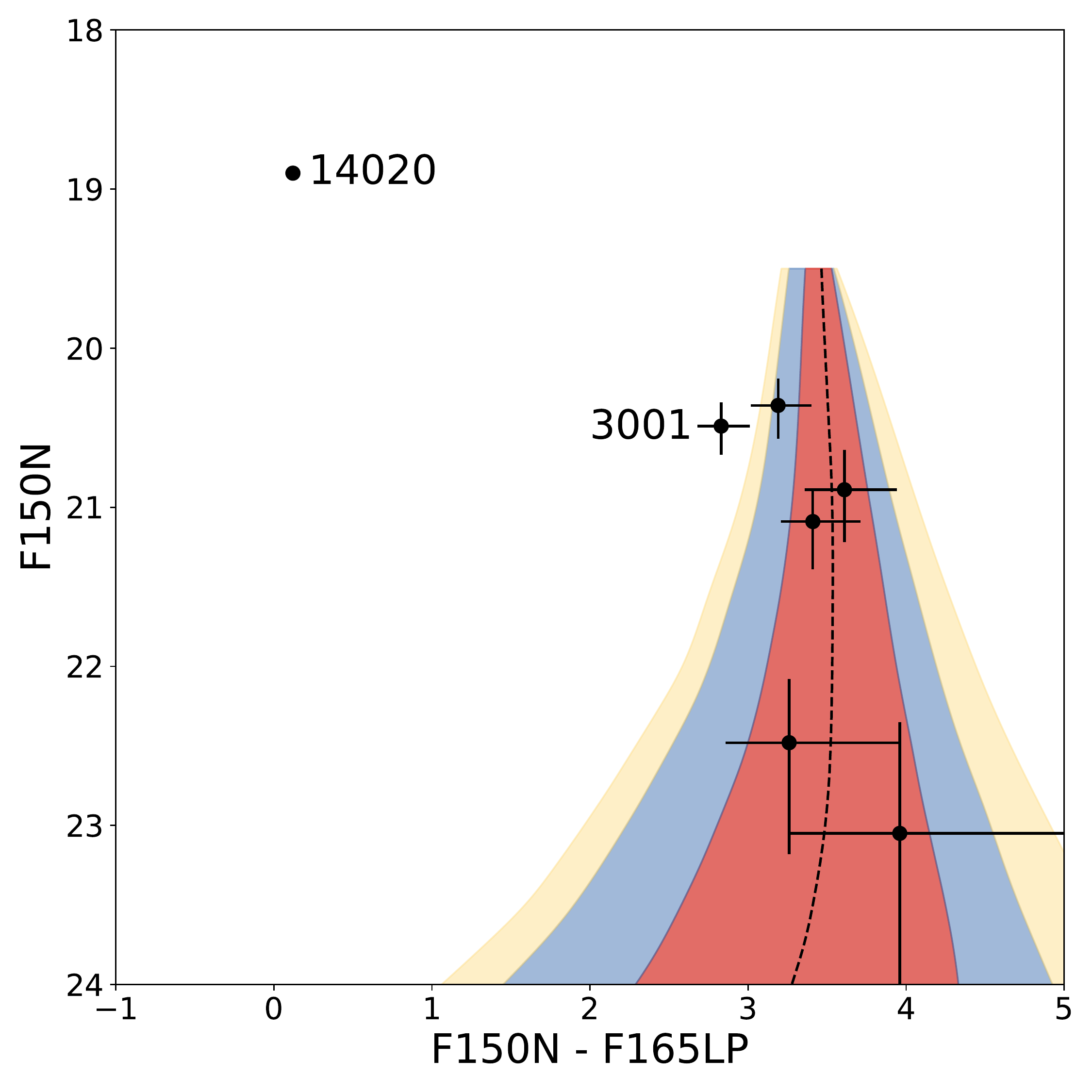}
    \caption{\footnotesize{FUV CMD of the binary BLs in M67, shown as dots with 1$\sigma$ error bars. The black dashed line traces the mean FUV color and magnitude of the M67 member stars (Section \ref{subsec:msmods}), while the red, blue, and yellow contours show the 1$\sigma$, 2$\sigma$, and 3$\sigma$ confidence intervals, respectively. The cutoff in the contours at approximately $\mathrm{F150N}=19.5$ is due to the exclusion of the BSSs. The two BLs with significant UV excess are indicated with their respective WOCS IDs.}}
    \label{fig:uvcmd}
\end{figure}

The black dashed line in Figure \ref{fig:uvcmd} is a cubic spline fit to the expected \textit{HST} FUV magnitudes and colors of M67 members computed in this process. We compute the expected count rates through each bandpass and use those rates to generate Poisson distributions that simulate observational noise. We sample the Poisson distribution 1000 times for each model atmosphere in order to compute confidence intervals, which we then also fit with cubic splines in order to create smooth contours. These intervals are shown in Figure \ref{fig:uvcmd}. 

We detect two BLs with significant ($>3\sigma$) excesses in their FUV flux compared to their single MS analogues: WOCS 3001 and WOCS 14020 (Figure \ref{fig:uvcmd}). 

\subsection{Blue Lurkers with Significant FUV Excesses}
\label{subsec:wd_detections}

The orbital periods of both WOCS 3001 and WOCS 14020 are $\lesssim$300~d, which is characteristic of Case B mass transfer from a red-giant-branch (RGB) donor that has not yet undergone the He flash (\citealt{Plavec1968, Ziol1970, Refsdal1971}), and which would leave behind the RGB core as a He WD. For example, \cite{Parsons2023} detected and measured the masses of He WDs in orbit around three field subgiants with masses about equal to the turnoff mass of M67. The longest orbital period of these three subgiant binaries is 461~d, similar to the orbital period of WOCS 14020. The authors determined these three systems to be the consequence of non-conservative Case B mass transfer, much like how both WOCS 14020 and WOCS 3001 may have formed (see Section \ref{subsec:mesa}). For specificity in our subsequent analyses, we therefore assume that the WD companions of both WOCS 3001 and WOCS 14020 are He WDs, albeit with unknown masses and gravities. 

In order to derive the effective temperatures of the WD companions, we first create UVBLUE model atmospheres of the BL primary stars corresponding to their effective temperatures derived in Section \ref{subsec:blmods}, and scaled to match their \textit{GALEX} NUV flux measurements. We then generate \cite{Koester2010} WD model atmospheres\footnote{We obtained the \cite{Koester2010} models from the Spanish Virtual Observatory Theoretical Model Services (\url{http://svo2.cab.inta-csic.es/theory/newov2/index.php?models=koester2}).}, scaled to an adopted distance to M67 of 859~pc (\citealt{CG2018}). We then interpolate the \cite{Koester2010} models to create a grid of WD model atmospheres that spans $7.0 \leq$~\logg~$\leq 7.6$ in steps of 0.1 and 9000~$\leq$~\teff~$\leq $~12,500~K in steps of 10~K, with each model scaled in radius according to the mass-radius-temperature relation described in \cite{Althaus2013}.

We add the scaled WD model spectra to the BL photospheric models in order to create composite spectra. We then convolve them with the \textit{HST} throughput curves (Section \ref{sec:obs}) with \texttt{pysynphot} to produce synthetic photometry for each bandpass. WOCS 14020 was also detected in the FUV by the \textit{GALEX} All-Sky Imaging Survey (\citealt{Bianchi2017}), and so we also convolve the composite spectrum with the \textit{GALEX} FUV bandpass. WOCS 3001 was not detected in the FUV by the \textit{GALEX} survey. 

Finally, we determine the best-fit WD temperature for each modeled \logg~ through $\chi^2$ minimization between all available measured and synthetic FUV photometry. In the case of WOCS 14020 we therefore simultaneously fit the \textit{HST} and \textit{GALEX} FUV measurements. For WOCS 3001, the F140N limit is not included in the minimization.

\begin{figure*}[t!]
    \centering
    \includegraphics[width=\linewidth]{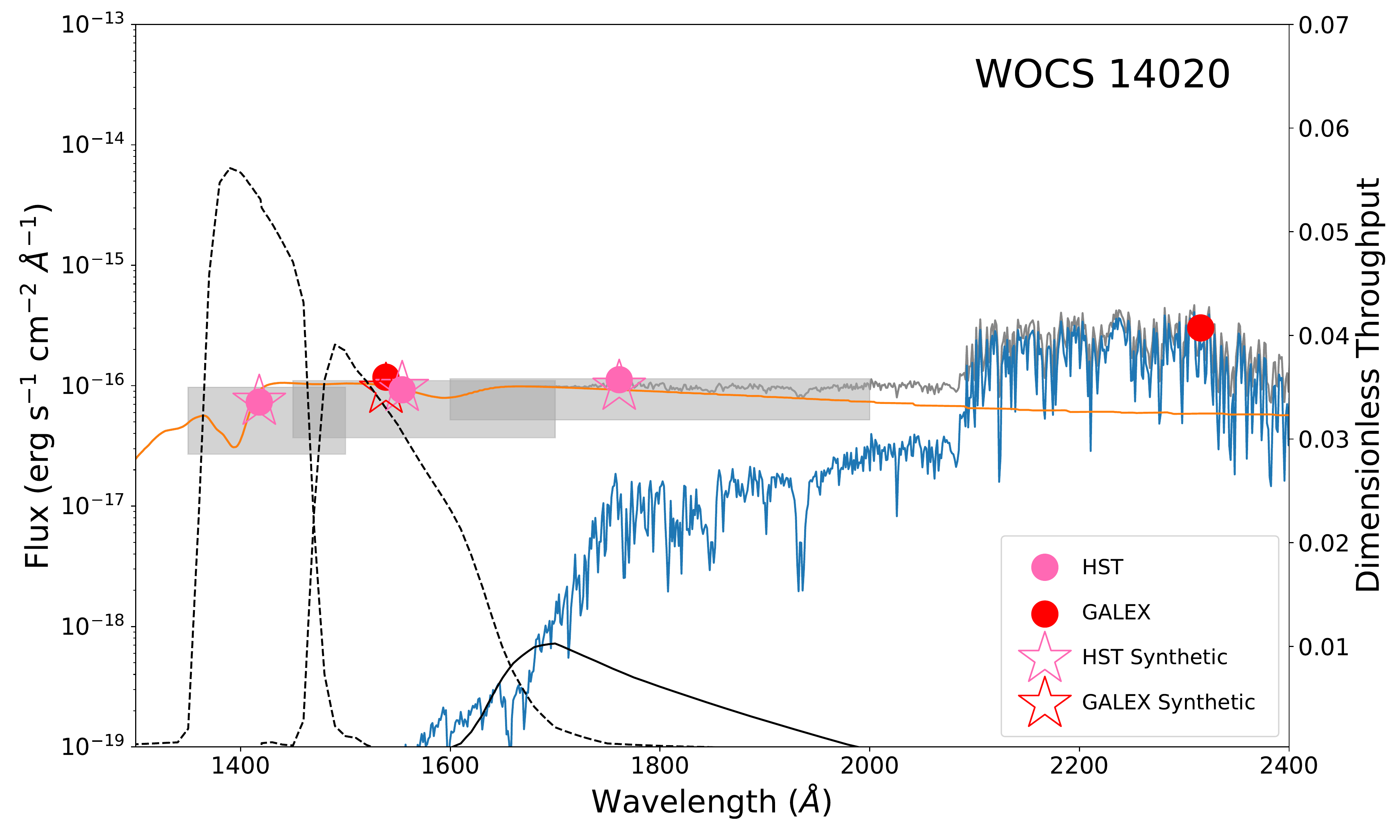}
    \caption{Composite SED of WOCS 14020. The SED of the BL primary is plotted in blue, and of the WD in orange. The composite spectrum is plotted in dark gray. The photometric measurements with \textit{HST} are shown with pink dots. The heights of the associated light gray boxes show the 1$\sigma$ uncertainties on the synthetic photometry due to the combined errors on the effective temperatures of both the BL and the WD. The widths of the light gray boxes represent the width of each \textit{HST} bandpass used in our analysis. The detailed shapes of the derived \textit{HST} narrowband filters F140N and F150N are shown with dashed black curves and of F165LP with a solid black curve; their relative throughputs are plotted with respect to the right-hand axis. We also include \textit{GALEX} measurements as red symbols. Synthetic photometry of the composite spectrum are shown with open star symbols in the respective color of their instruments, and show close agreement with all measurements.}
    \label{fig:bl_sed_14020}
\end{figure*}

The best-fit WD temperatures are presented in Table \ref{blue:tab:wdtab} as a function of \logg. The uncertainties listed on each determined temperature are the formal uncertainties resulting from $\chi^2$ minimization. We also include in Table \ref{blue:tab:wdtab} estimates of the WD cooling ages adopted from \cite{Althaus2013} for the same WD temperatures and \logg. 

Based on the binary mass function of WOCS 3001 (Table \ref{blue:tab:summ}), we are able to constrain the WD companion mass to between 0.35--0.45~\Msolar\, assuming a primary mass of 1.3~\Msolar\, (see Section \ref{subsec:mesa}), where the upper bound is the maximum mass of a He WD (\citealt{Hayashi1962, Iben1991}). The binary mass function of WOCS 14020 permits a wider range of WD companion masses, and we have fit WD models in the range 0.25--0.45~\Msolar, assuming a primary mass of 1~\Msolar\, (Section \ref{subsec:mesa}).

\begin{figure*}[htbp]
    \centering
    \gridline{\fig{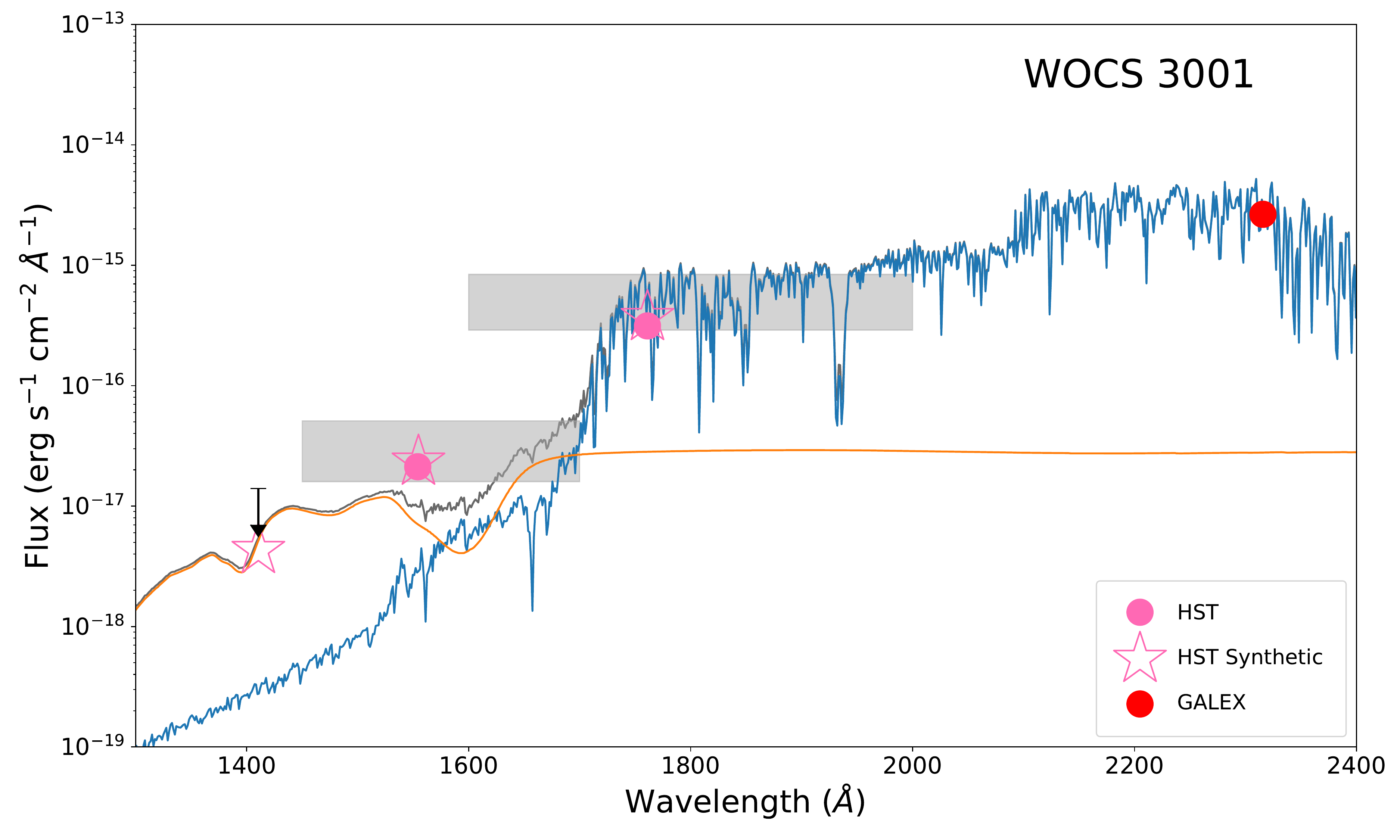}{0.5\linewidth}{}
    \fig{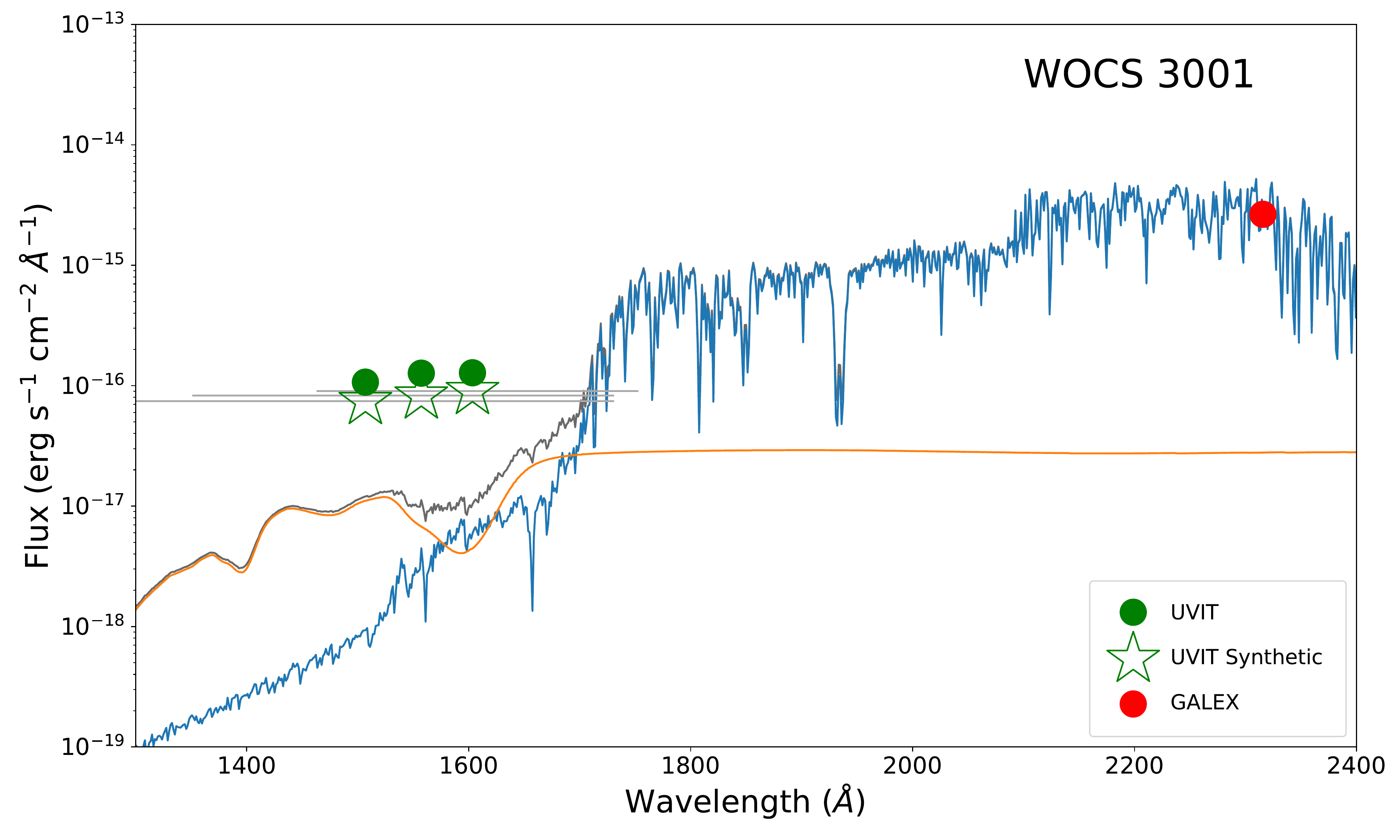}{0.5\linewidth}{}}
    \caption{Left panel: Similar to Figure \ref{fig:bl_sed_14020}, but for WOCS 3001. Since WOCS 3001 was not detected in F140N, we show the 3$\sigma$ upper limit on the flux in that bandpass. We again show the synthetic F140N flux of the composite spectrum as computed by \texttt{pysynphot} with an open star symbol. Right panel: Published photometric measurements with UVIT are shown with green dots. The associated bandwidths are shown with horizontal lines. Synthetic photometry of the composite spectrum using the UVIT throughput curves are shown with open star symbols. The UVIT measurements are significantly higher than \textit{HST} measurements and the composite spectrum, but their agreement with the synthetic photometry is reasonable due to the broad bandpasses capturing some of the Wien tail from the BL primary.}
    \label{fig:bl_sed_3001}
\end{figure*}

We show representative best-fit composite (BL+WD) spectra for WOCS 14020 and WOCS 3001 in Figures \ref{fig:bl_sed_14020} and \ref{fig:bl_sed_3001}, respectively. In the case of WOCS 14020 we show a WD model atmosphere with \logg~$=7.4$ and an effective temperature of 11,800~K (corresponding to a mass of 0.37~\Msolar), and for WOCS 3001 we show the case of \logg~$=7.5$ and an effective temperature of 10,400~K (corresponding to a mass of 0.39~\Msolar). We stress again that the precise masses of the WD companions are as yet unknown.

In both figures the UVBLUE model atmosphere of the BL primary is plotted in blue, the WD model atmosphere is plotted in orange, and the composite spectrum is plotted in gray. Following \cite{Gosnell2015}, the widths of the gray boxes correspond to the widths of the \textit{HST} bandpasses. The heights of the gray boxes correspond to the 1$\sigma$ uncertainty on the synthetic photometry as a consequence of the uncertainties of the effective temperatures of both the BL primary (Table \ref{blue:tab:summ}) and the WD companion (Table \ref{blue:tab:wdtab}), where we take the uncertainties on the WD temperatures to be the range of best-fit temperatures.

Finally, WOCS 3001 also was observed by \cite{Jadhav2019}, who measured a significant excess in FUV flux using the F148W, F154W, and F169M bandpasses of the \textit{Ultra-Violet Imaging Telescope} (UVIT) on board AstroSat (\citealt{Agrawal2006, Kumar2012}). Their measured fluxes are substantially higher than our \textit{HST} F150N detection and F140N upper limit (Figure \ref{fig:bl_sed_3001}). 

We have convolved our composite spectrum for WOCS 3001 with the FUV bandpasses of UVIT as described in \cite{Tandon2017}. The resulting synthetic photometry (Figure \ref{fig:bl_sed_3001}) is similarly higher than the \textit{HST} fluxes, due to broader UVIT bandpasses capturing some of the Wien tail from the BL primary. The UVIT synthetic photometry of our model is slightly lower than the UVIT photometric measurements.

\cite{Jadhav2019} determined a photometric temperature of $12500\pm250$~K for the WD companion and a mass of between 0.30--0.45~\Msolar. Our determined photometric temperatures are significantly lower (Table \ref{blue:tab:wdtab}) due to the different flux measurements and differing assumptions about the radius and \logg~ of the WD.

\centerwidetable
\begin{deluxetable*}{cccccc}
\tabletypesize{\footnotesize}
\tablecaption{He WD Parameters\label{blue:tab:wdtab}}
\tablehead{\colhead{WOCS ID} & \colhead{Modeled \logg\tablenotemark{a}} & \colhead{Best-Fit WD Temperature\tablenotemark{a}}  & \colhead{WD Mass\tablenotemark{b}} & \colhead{WD Radius\tablenotemark{b}} & \colhead{Cooling Age\tablenotemark{b}} \\
\colhead{} & \colhead{} & \colhead{(K)} & \colhead{(\Msolar)} & \colhead{(\Rsolar)} & \colhead{(Myr)}}
\tablewidth{0pt}
\startdata
14020 & 7.0 & $11040^{+20}_{-30}$ & 0.26 & 0.027 & $350\pm8$ \\
& 7.1 & $11230^{+20}_{-30}$ & 0.28 & 0.025 & $290\pm18$ \\
& 7.2 & $11400^{+30}_{-30}$ & 0.31 & 0.023 & $400\pm16$ \\
& 7.3 & $11590^{+30}_{-30}$ & 0.34 & 0.022 & $450\pm10$ \\
& 7.4 & $11800^{+40}_{-30}$ & 0.37 & 0.020 & $400\pm7$ \\
& 7.5 & $12010^{+30}_{-40}$ & 0.40 & 0.019 & $500\pm15$ \\
& 7.6 & $12220^{+40}_{-40}$ & 0.44 & 0.017 & $540\pm5$ \\
\\
3001 & 7.4 & $10280^{+50}_{-50}$ & 0.36 & 0.020 & $630\pm16$ \\
& 7.5 & $10400^{+50}_{-60}$ & 0.39 & 0.019 & $720\pm20$ \\
& 7.6 & $10490^{+50}_{-70}$ & 0.43 & 0.017 & $900\pm10$ \\
\enddata
\tablenotetext{a}{Interpolated from the \cite{Koester2010} WD model atmospheres}
\tablenotetext{b}{Derived from the WD mass-radius-temperature relation described in \cite{Althaus2013}}
\end{deluxetable*}


\section{Discussion}
\label{sec:discuss}

\subsection{Spin-down of the Blue Lurkers}
\label{subsec:spin_down}

\begin{figure*}[t!]
    \centering
    \includegraphics[width=\linewidth]{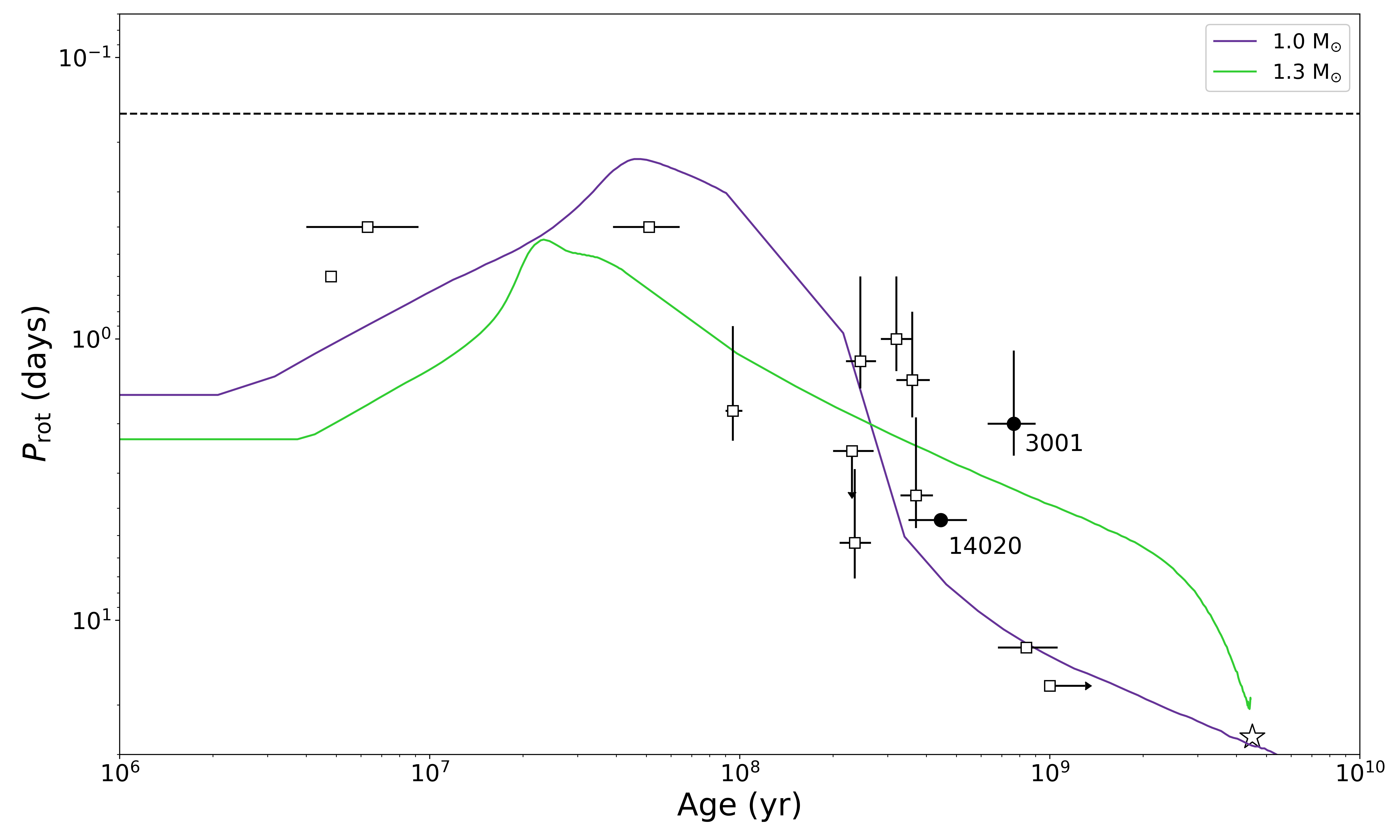}
    \caption{Age-rotation diagram of the two BLs with detected WD companions. The filled points correspond to the estimated cooling ages and measured rotation periods of the BLs with detected WD companions, and 1$\sigma$ uncertainties are shown with black bars. The open squares and associated error bars correspond to BSSs in NGC 188 with hot WD companions (ages~$\gtrsim 10^8$~yr; \citealt{Gosnell2015, Gosnell2019}) and other known mass-transfer products as described in Table 1 of \cite{Leiner2018} and references therein. The curves trace spin evolution models for a 1.0 and 1.3~\Msolar~single star (shown in purple and green, respectively) from \cite{Amard2019}. The dashed black line shows the critical rotation period for a 1~\Msolar, 1~\Rsolar~star from \cite{Ekstrom2008}. The age and rotation period of the Sun are shown with an open star symbol for reference.}
    \label{fig:prot_gyro}
\end{figure*}

\cite{Leiner2018} showed that the rotation periods of known BSSs with WD cooling ages fall on expected rotational evolution model tracks for single MS FGK-type stars. The rotation periods and cooling ages of the two BLs with WDs follow the same trend (Figure \ref{fig:prot_gyro}). The agreement between the BLs and the BSSs corroborates the cooling ages described in Table \ref{blue:tab:wdtab}. It also suggests that the BLs exhibit the same physics of spin evolution as do the BSSs. This implies that the BLs may have had a similar history of spin-up as a consequence of mass transfer followed by spin-down through magnetic braking, evolving in parallel with the BSSs.

Specifically, in Figure \ref{fig:prot_gyro} we compare the observed rotation rates and estimated WD cooling ages of the BLs to the rotational evolution models of \cite{Amard2019}. We choose the 1.0~\Msolar\, and 1.3~\Msolar\, solar-metallicity, fast-rotating models based on the estimated masses of the BL primary stars (see Section \ref{subsec:mesa}). In brief, the models consider angular momentum evolution through disk-star interactions on the pre-main-sequence (\citealt{Ghosh1979, Bouvier2014}), the transport of angular momentum between the core and the envelope (\citealt{Mathis2004, Decressin2009, Mathis2018}), and the loss of angular momentum through stellar winds (\citealt{Schatz1962, Kawaler1988, Matt2015}). 

In the \cite{Amard2019} models, stars begin their evolution tidally locked to a circumstellar disk, maintaining constant rotation for the first few $\times10^6$~yr. The most rapidly rotating stars with masses $<1.2$~\Msolar\, have initial rotation periods of 1.6~d, consistent with the fastest rotating stars in young clusters (\citealt{Gallet2015}). After the stars decouple from their disks, they contract and spin up to the MS phase, achieving rotation periods of a significant fraction of the critical velocity. More massive stars have minimum initial rotation periods of 2.3~d in order to avoid supercritical rotation during the contraction phase. After stars arrive on the MS, angular momentum evolves through core-envelope interactions and magnetized winds.

The rotation periods of the BLs are consistent with the \cite{Amard2019} spin evolution models. The more rapid rotation of WOCS 3001 at a later age may be explained in the 1.3~\Msolar\, model by the absence of a deep convective envelope given the higher BL effective temperature and the consequent lack of the strong magnetic fields necessary for efficient magnetic braking (\citealt{Matt2012, Garraffo2018}). 

As \cite{Sun2021} demonstrated, once mass transfer has ceased a BSS can have the same interior structure as a single star in the same location on the HR diagram. The same may be true of the BLs, and they are structurally similar to single stars of the same mass. If this similar structure leads to similar magnetic fields and winds, their rotation may evolve in a similar manner.

\subsection{The White Dwarf Detection Rate}
\label{subsec:detect}

Out of our sample of eight binary BLs we detected two BLs with FUV excesses indicative of WD companions, which sets a lower limit of $25\pm18$\% on the mass-transfer formation frequency of the BLs in our sample and $18\pm13$\% of the total BL population described in L19. This is similar to the lower limit placed by \cite{Pandey2021} on the mass-transfer formation rate among the classical BSSs in M67, which the authors determined to be $36\pm16$\% (our Poisson uncertainty).

It is likely that mass transfer was responsible for the creation of a higher fraction of the BLs, the remainder of which have cooled below our detection limit. Including WOCS 3001 and WOCS 14020, there are five binary BLs in the sample with binary mass functions compatible with WD companions (see Table~\ref{blue:tab:summ}). They all have minimum secondary masses of 0.15--0.5~\Msolar\, assuming as did L19 that the primary masses range between 0.9--1.3~\Msolar. Two of the eight binary BLs have mass functions too large to be compatible with WD secondaries (WOCS 2068, minimum companion mass of $\sim$0.7~\Msolar, and WOCS 6025, minimum companion mass of $\sim$1.1~\Msolar). The final system (WOCS 11006) does not yet have a complete orbital solution and so the binary mass function is unknown. 

Two of the binaries with companion lower mass limits $\lesssim$0.5~\Msolar\, are WOCS 3001 and WOCS 14020. If the three additional BLs with low companion mass limits were also formed through mass transfer, the mass-transfer formation frequency becomes $45\pm20$\% considering the full population of 11 BLs. If the orbit solution for the binary WOCS 11006 has a mass function consistent with mass transfer, then the mass-transfer formation frequency becomes $55\pm27$\%. For comparison, \cite{Gosnell2015} estimated a mass-transfer formation frequency of 67\% for the BSSs of NGC 188 based on WD detections there.

Assuming a donor mass of between 1.3 and 1.5~\Msolar, typical orbital periods at the onset of Case B mass transfer range from a few to a few hundred days (\citealt{Thomas1977}). The final orbital period after mass transfer may be estimated through the relation of \cite{Rappaport1995}. Assuming a maximum possible He WD mass of approximately 0.45~\Msolar, the maximum period resulting from Case B mass transfer in M67 is $P_{\mathrm{orb}}\approx800$~d.

Building on L19, we note that among the 5 binaries with low companion mass limits, 4 have orbital periods of less than 800 days. (Or 4 out of 6 if WOCS 11006 is included.) WD companions are now detected for 2 of these 4, confirming their mass-transfer origins. While the numbers are small, there is a suggestion in these data that mass transfer from RGB stars preferentially forms BLs, whereas mass transfer from an asymptotic-giant-branch (AGB) companion has been proposed as a more common formation channel for classical BSSs in NGC 188 and M67, which predominantly have orbital periods $\gtrsim1000$ days (\citealt{Geller2011, Latham2007}). 

\begin{figure}[t!]
    \centering
    \includegraphics[width=\linewidth]{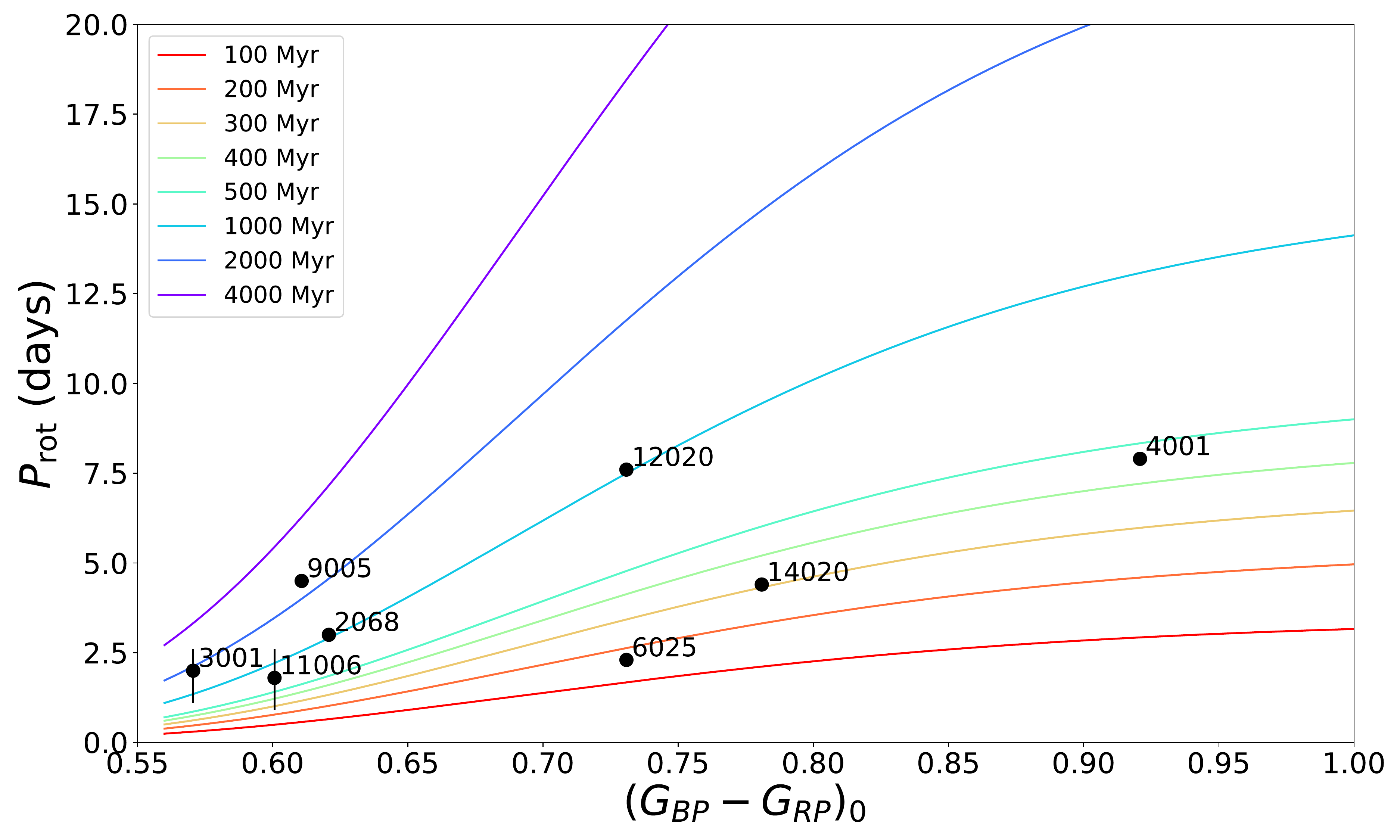}
    \caption{Color-rotation plot of the eight binary BLs compared to the gyrochronology models of \cite{Angus2019}; compare with Figure 1 of L19. We use the photometry of \textit{Gaia} DR2, dereddened using $E(B-V) = 0.041 \pm 0.004$ (\citealt{Taylor2007}) and following the procedure of \cite{GaiaCollab2018a}.}
    \label{fig:prot_color}
\end{figure}

Based on gyrochronology, L19 argue from their rapid rotation that all of the BLs have ages of less than 1 Gyr. We are able to detect He WD companions with cooling ages of up to approximately 900 Myr. Given these, we would expect to detect $3.6\pm1.9$ He WDs among these four BLs. Considering the updated gyrochronology models of \cite{Angus2019}, as shown in Figure \ref{fig:prot_color}, it is also possible that the eight binary BLs formed within the last $\sim$2 Gyr. In that case, we would expect to detect $1.8\pm1.3$ He WDs. Our detection of two WDs in either case is therefore reasonable. We also note that for the two longer-period ($P_{\mathrm{orb}}>1000$~d) binaries for which mass transfer from an AGB companion is more probable (\citealt{Kippenhahn1967, Pacz1971}), the non-detection of WD companions does not rule out mass-transfer formation since the maximum detectable cooling age of a CO WD using our methodology is approximately 400~Myr (\citealt{Gosnell2015}).

\cite{Geller2021} detected 40 single-lined binaries within the upper MS of M67 with $P_{\mathrm{orb}}<800$~d, from which these 4 binaries are drawn. If as an upper bound we presume that BL formation among solar-type stars has been constant over the 4~Gyr cluster lifetime, given that the 4 BLs discussed here could all have been formed within the last 1--2~Gyr there might be of order 8--16 BLs with He WD companions among these 40 binaries, the majority of which have slowed to intermediate rotation periods as L19 suggest. Similar numbers of BLs have been found in \textit{N}-body simulations of open clusters (\citealt{PZ2001, Geller2013}). \cite{Geller2021} also detected 15 double-lined spectroscopic binaries on the upper MS with $P_{\mathrm{orb}}<100$~d, some of which may evolve into BLs in the future.

In summary, with the detection of 2 WDs among 8 BLs, we conclude that mass transfer is viable as the primary mechanism to create binary BLs, and as such the BLs within the M67 upper MS represent an extension of the BSS population to lower luminosities. 

\subsection{Modeling Mass-Transfer Histories}
\label{subsec:mesa}

We have assumed that the WD companions are He WDs since the orbital periods are indicative of Case B mass transfer from an RGB companion. As L19 note, the BLs pose a challenge to mass-transfer theory because it has been widely predicted that, given the most probable initial mass ratios of the systems that created the BL+WD binaries, RGB mass transfer would become unstable and result in common envelope evolution (see for example \citealt{Webbink1988} and \citealt{Chen2008}). As an example of mass transfer that occurred under such circumstances, \cite{Sun2021} studied the BSS WOCS 5379 in NGC 188 with the \texttt{Modules for Experiments in Stellar Astrophysics}\footnote{\url{https://docs.mesastar.org/en/latest/index.html}} (MESA; \citealt{Paxton2011, Paxton2013, Paxton2015, Paxton2018, Paxton2019}). The authors found that WOCS 5379 could have formed through highly non-conservative Case B mass transfer. The mass transfer begins with a short phase of rapid and stable mass transfer, followed by a longer period of slower stable mass transfer. In this section we explore the possibility that the BLs in M67 may have formed in the same way.

 We adapt the MESA version r15140\footnote{Compiled with MESA SDK version 21.11.1 for Mac OS (\citealt{Townsend2021})} binary test case \texttt{evolve\_both\_stars} using standard assumptions about stellar and binary evolution, including the mass transfer prescriptions of \cite{Eggleton1983} and \cite{Ritter1988} as well as the RGB wind prescription of \cite{Reimers1975} with a scaling factor of 0.5. Our inlists and starting models are available at \url{https://zenodo.org/record/7502526}. We do not present these models as the definitive histories of WOCS 14020 and WOCS 3001, but as initial plausibility arguments for the proposed mass-transfer formation of these systems.

\begin{figure}[t]
    \centering
    \includegraphics[width=\linewidth]{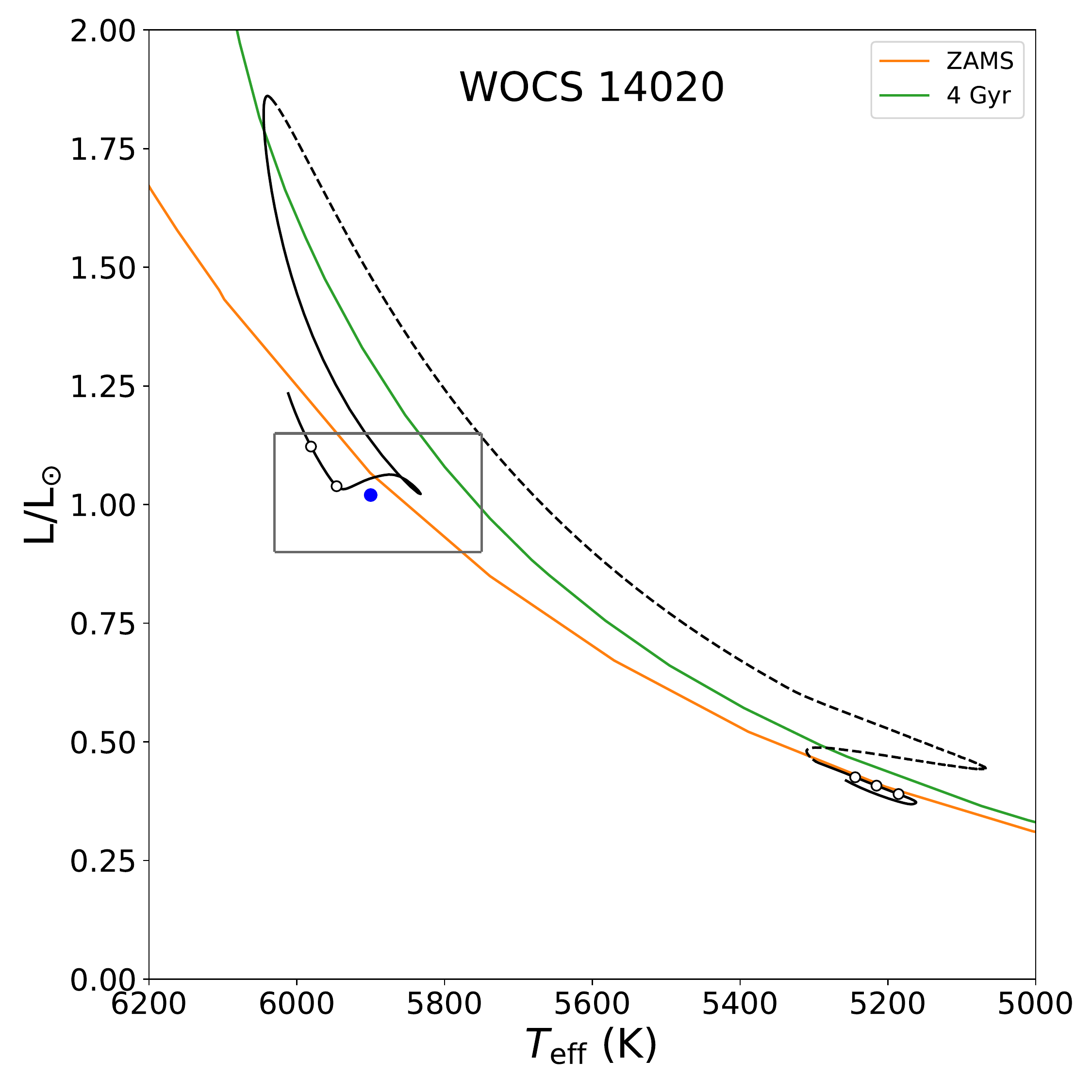}
    \caption{Evolutionary track of WOCS 14020 on the HR diagram. The dashed black line traces the evolution of WOCS 14020 during mass transfer, while the solid black line traces the history before and after active mass transfer. Open circles represent 1 Gyr timesteps. The blue dot represents the observed properties of WOCS 14020, and the gray box represents their 1$\sigma$ uncertainties. The orange and green lines represent the zero-age MS and 4 Gyr PARSEC (\citealt{Bressan2012}) isochrones, respectively.}
    \label{fig:14020_evol_track}
\end{figure}

Our initial configuration for an evolutionary model of WOCS 14020 is a 1.38~\Msolar\, donor star and a 0.87~\Msolar\, accretor star (corresponding to a mass ratio of $M_d/M_a \approx 1.6$) in a 120~d circular orbit. The donor star evolves off the MS first and ascends the RGB; at the point that mass transfer begins the donor mass has been reduced to 1.34~\Msolar\, by stellar winds. Mass transfer then proceeds non-conservatively with 17\% of the mass lost from the primary being accreted by the secondary, the other 83\% being lost by the donor as a fast wind through the $\alpha$ mechanism as defined by \cite{Tauris2006}. The resulting system is a 1.04~\Msolar\, BL and a 0.36~\Msolar\, He WD with \logg~$\approx7.3$ in a 362~d orbit.

We present in Figure \ref{fig:14020_evol_track} the evolutionary track of the accretor star in comparison to the observed properties of WOCS 14020.  There is excellent agreement between the evolutionary track at an age of 4~Gyr and the current observed properties of the current WOCS 14020. This shows that it is possible to reproduce the observed BL system through non-conservative mass transfer. 

The evolution of WOCS 14020 follows closely that of WOCS 5379 (\citealt{Sun2021}), to which the reader is referred for more detail. In short, mass transfer in the case of WOCS 14020 also begins with a brief phase of rapid and stable mass transfer followed by a longer period of slower stable mass transfer. WOCS 14020 also exhibits the same pattern of initial reduction in \teff\, and subsequent increase as seen in the evolution of WOCS 5379.

\begin{figure}[t]
    \centering
    \includegraphics[width=\linewidth]{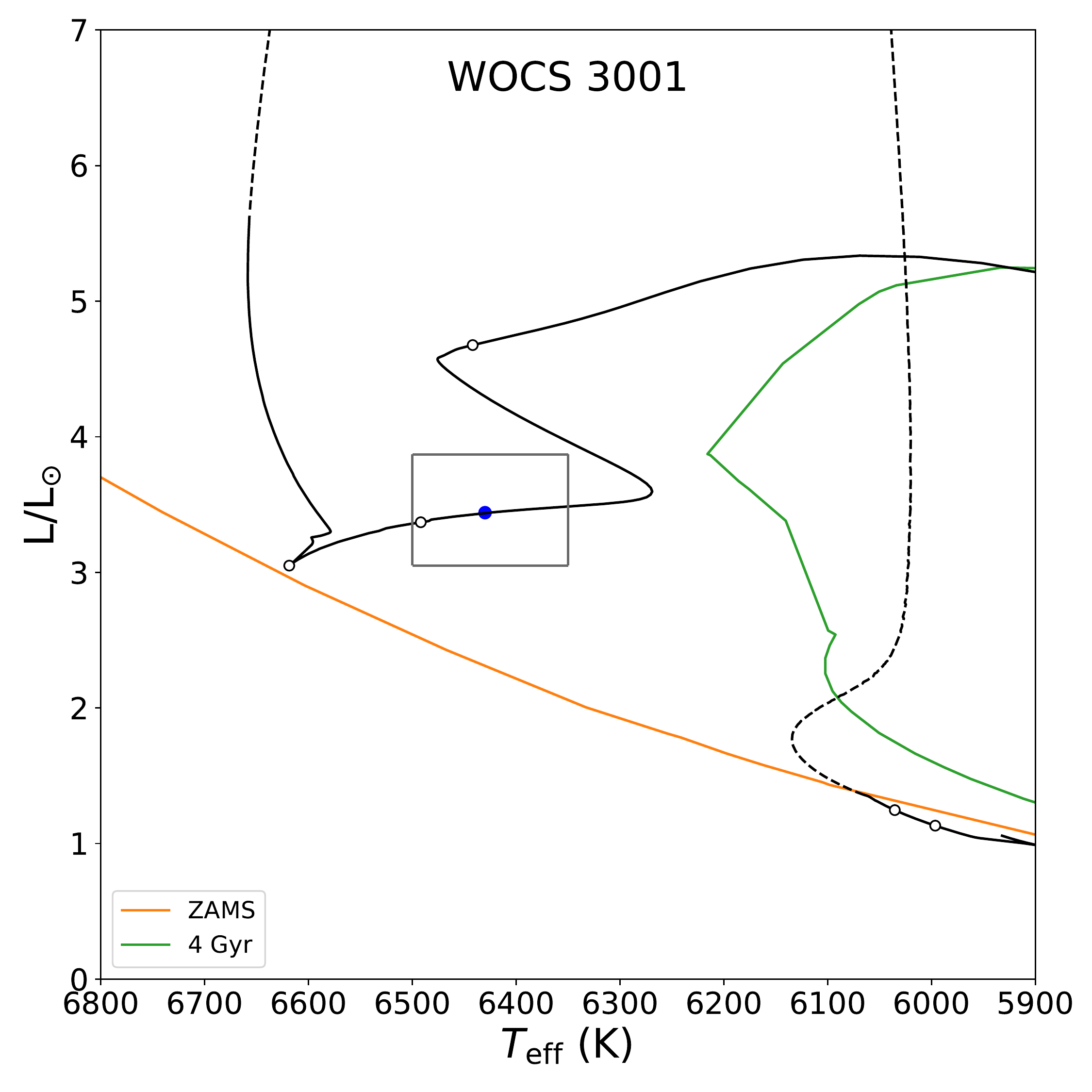}
    \caption{Evolutionary track of WOCS 3001 on the HR diagram. The lines are the same as in Figure \ref{fig:14020_evol_track}.}
    \label{fig:3001_evol_track}
\end{figure}

In the case of WOCS 3001, the best-fit initial configuration is a 1.49~\Msolar\, donor star and a 1.08~\Msolar\, accretor star ($M_d/M_a \approx 1.4$) in a 34.5~d circular orbit. Because of the relatively close orbit, mass transfer begins before the donor star loses an appreciable amount of its mass to winds. Mass transfer in this case proceeds with an efficiency of 18\%, again through the $\alpha$ mechanism. The resulting system is a 1.29~\Msolar\, BL and a 0.32~\Msolar\, He WD with \logg~$\approx7.2$ in a 121~d orbit. The evolutionary track of WOCS 3001 is shown in Figure \ref{fig:3001_evol_track}. The evolutionary track of WOCS 3001 lies within the ranges of uncertainty at an age of 4~Gyr, again suggesting that a mass-transfer origin is a plausible hypothesis for its formation.

Future detailed modeling work is needed in order to comprehensively investigate the parameter space. The models presented here, however, suggest that non-conservative Case B mass transfer as \cite{Sun2021} proposed for the formation of WOCS 5379 may be common.

\section{Summary}
\label{sec:summary}

In this work we conduct a FUV photometric survey of the eight binary BLs in M67 with \textit{HST} ACS/SBC. We find evidence for two WD companions, one with \teff~$=$~11,000--12,200~K and the other with \teff~$=$~10,300--10,500~K. These photometric temperatures correspond to cooling ages of $\sim$300--540~Myr and $\sim$600--900~Myr, respectively. The orbital properties of these BLs suggest that they are He WDs resulting from Case B mass transfer, though follow-up FUV spectroscopy is necessary in order to confirm this.

We find that the observed rotation periods and cooling ages of the BLs with detected WD companions are consistent with spin-up during MT and models of subsequent single-star spin-down, similar to as found for classical BSSs. 

Our detection of two WDs is consistent with mass transfer being the primary formation mechanism for the BLs. Based on the orbits of the observed BL binary population, we determine that the BL mass-transfer formation frequency could be as high as 55$\pm$27\%, similar to the 67\% determined for the classical BSSs in the older open cluster NGC 188. 

Together, these results further support the hypothesis that the BLs are lower-luminosity analogues to the classical BSSs, and that together these populations represent a continuum of products of solar-type binary evolution. There are likely many more BLs waiting to be discovered.

Both of our WD detections are for binaries with shorter orbital periods suggesting an RGB mass transfer origin. If we assume the two BLs with detected WDs both have He WD companions, we are able to successfully reproduce the observed properties of both systems through models of highly non-conservative Case B mass transfer. If He WDs can be confirmed in these systems (e.g., with UV spectroscopic observations), these BLs suggest that stable non-conservative Case B mass transfer is more common than has been previously supposed. The examples of such mass transfer presented in this work may serve to inform future theoretical study in this regard.

Among the eight binary BLs in our sample, four have shorter orbital periods indicative of RGB mass transfer and one has an orbital period indicative of AGB mass transfer. (Two have mass functions or SEDs that are not suggestive of a mass transfer origin, and one has a long orbital period but no orbital solution yet.) In contrast, most of the classical BSSs in M67, and in the similarly old cluster NGC 188, have longer orbital periods suggestive of frequent AGB mass transfer. It is possible that RGB mass transfer origins are more common amongst BLs than BSSs. This could result if AGB mass transfer tends to be more conservative than RGB mass transfer, or if stable AGB mass transfer favors more massive accretor stars. Larger samples are necessary in order to securely establish the relative rates and the physics of mass-transfer formation pathways among BLs relative to BSSs.

With the ongoing \textit{TESS} and future \textit{PLATO} missions, there will doubtless be many more BLs discovered both in clusters and in the field. The BLs of M67 will be foundational for future studies of these systems in other environments and in our understanding of the limits of mass transfer.

\vspace{1cm}

The University of Wisconsin-Madison authors acknowledge funding support from NSF AST-1714506 and the Wisconsin Alumni Research Foundation. ACN acknowledges funding support from Sigma Xi grant G20201001111059492. NMG is a Cottrell Scholar receiving support from the Research Corporation for Science Advancement under grant ID 27528. EML is supported by an NSF Astronomy and Astrophysics Postdoctoral Fellowship under award AST-1801937. The authors also gratefully acknowledge the many Wisconsin undergraduate and graduate students who have contributed to the WIYN Open Cluster Study radial-velocity database.

This research is based on observations made with the NASA/ESA \textit{Hubble Space Telescope} obtained from the Space Telescope Science Institute, which is operated by the Association of Universities for Research in Astronomy, Inc., under NASA contract NAS 5–26555. These observations are associated with program 16244. All of the \textit{HST} data used in this paper can be found in MAST: \dataset[10.17909/y2sh-s027]{http://dx.doi.org/10.17909/y2sh-s027}

This work has made use of data from the European Space Agency (ESA) mission {\it Gaia} (\url{https://www.cosmos.esa.int/gaia}), processed by the {\it Gaia} Data Processing and Analysis Consortium (DPAC, \url{https://www.cosmos.esa.int/web/gaia/dpac/consortium}). Funding for the DPAC has been provided by national institutions, in particular the institutions participating in the {\it Gaia} Multilateral Agreement.

This work made use of the Third Data Release of the GALAH Survey (\citealt{Buder2021}). The GALAH Survey is based on data acquired through the Australian Astronomical Observatory, under programs: A/2013B/13 (The GALAH pilot survey); A/2014A/25, A/2015A/19, A2017A/18 (The GALAH survey phase 1); A2018A/18 (Open clusters with HERMES); A2019A/1 (Hierarchical star formation in Ori OB1); A2019A/15 (The GALAH survey phase 2); A/2015B/19, A/2016A/22, A/2016B/10, A/2017B/16, A/2018B/15 (The HERMES-TESS program); and A/2015A/3, A/2015B/1, A/2015B/19, A/2016A/22, A/2016B/12, A/2017A/14 (The HERMES K2-follow-up program). We acknowledge the traditional owners of the land on which the AAT stands, the Gamilaraay people, and pay our respects to elders past and present. This paper includes data that has been provided by AAO Data Central (\url{https://datacentral.org.au/}).

This research is based partially on results obtained from the AstroSat mission of the Indian Space Research Organisation (ISRO) archived at the Indian Space Science Data Centre (ISSDC).

This research has made use of the Spanish Virtual Observatory (\url{https://svo.cab.inta-csic.es}) project funded by MCIN/AEI/10.13039/501100011033/ through grant PID2020-112949GB-I00.

This work has made use of Astropy (\url{http://www.astropy.org}), a community-developed core Python package for Astronomy \citep{astropy2013, astropy2018, astropy2022}. This work has also made use of PyAstronomy (\citealt{Czesla2019}).

This work was conducted at the University of Wisconsin-Madison which is located on occupied ancestral land of the Ho-chunk Nation, and observations for this work were conducted on the traditional lands of the Tohono O'odham Nation. We respect the inherent sovereignty of these nations, along with the other eleven First Nations in Wisconsin as well as the southwestern tribes of Arizona. We honor with gratitude these lands and the peoples who have stewarded them, and who continue to steward them, throughout the generations.

\vspace{1cm}
\bibliography{blue.bib}{}
\bibliographystyle{aasjournal}

\end{document}